\documentclass[pra,showpacs,twocolumn,superscriptaddress,longbibliography]{revtex4-2}
\usepackage{amsmath}   
\usepackage{amssymb}
\usepackage{graphicx}
\usepackage{dcolumn}
\usepackage{bm}
\usepackage{amsmath}
\usepackage{graphicx}
\usepackage{amsfonts}
\usepackage{subfigure}
\usepackage{graphicx}
\usepackage{color}
\usepackage{xcolor}
\usepackage{soul}
\usepackage[normalem]{ulem}
\usepackage{braket}
\usepackage{amssymb}
\usepackage{hyperref}
\usepackage{comment}
\usepackage{xr}
\usepackage{bbold}
\usepackage[T1]{fontenc}

\usepackage{titlesec}
\hypersetup{
    colorlinks=true,
    citecolor=blue,
    linkcolor=blue,
    filecolor=black,      
    urlcolor=blue,
}
\begin{document}
\title{Near-deterministic single-atom loading on a photonic integrated circuit}
\author{Xinchao Zhou}
\altaffiliation{Present address: Center for Hybrid Quantum Networks (Hy-Q), Niels Bohr Institute,
University of Copenhagen, Jagtvej 155A, Copenhagen DK-2200, Denmark}
\affiliation{Department of Physics and Astronomy, Purdue University, West Lafayette, IN 47907, USA}
\author{Ahreum Lee}
\affiliation{Department of Physics and Astronomy, Purdue University, West Lafayette, IN 47907, USA}
\author{Dipanjan Das}
\affiliation{Department of Physics and Astronomy, Purdue University, West Lafayette, IN 47907, USA}
\author{Saivirinchi Prabandhakavi}
\affiliation{Department of Physics and Astronomy, Purdue University, West Lafayette, IN 47907, USA}
\author{Chen-Lung Hung}
\email{clhung@purdue.edu}
\affiliation{Department of Physics and Astronomy, Purdue University, West Lafayette, IN 47907, USA}
\affiliation{Purdue Quantum Science and Engineering Institute, Purdue University, West Lafayette, IN 47907, USA}
\date{\today}

\begin{abstract}
Coupling identical quantum emitters to a photonic integrated circuit (PIC) is a key step for scaling up emitter-photon interfaces for quantum science and information processing. Neutral atoms are attractive candidates due to their indistinguishability and controllability. However, experimental realizations of efficient atom trapping on a PIC while achieving strong single atom-photon coupling has so-far remained elusive. Here, we demonstrate near-deterministic single-atom loading on a microring resonator circuit, reaching single-atom cooperativity parameter $C>1$ for strong coupling in cavity quantum electrodynamics. We utilize a precision optical conveyor belt, formed by a moving optical lattice in an optical tweezer, to steadily deliver trapped atoms onto a PIC. By continuously monitoring the transmission of probe photons through the circuit, which is sensitive to the proximity of single atoms near a microring resonator, we detect mean occupancy of $1.5$ from 70 occupied lattice sites in a conveyor-belt transport of 4~nm position reproducibility. Based upon real-time feedback, we deterministically transfer the delivered atoms into a stationary trap on the microring, achieving $82\%$ ($18\%$) probability of single-(two-)atom transfer. Our technique can be extended to deterministic, highly efficient atom array assembly, providing a scalable route for neutral atom integration with PICs of complex functionalities.
\end{abstract}

\maketitle

\section{Introduction}

In recent years, photonic integrated circuits (PICs) have emerged as a highly scalable platform for manipulating visible and near-infrared photons~\cite{wang2020integrated, luo2023recent, 2023NaturePhotonics_laserchip,2026Nature_ChenFiberLikeLoss, 2021NaturePhotonics_laser,2024NatPhotonics_review}. However, inducing the quantum nonlinearities required for high-fidelity quantum information processing~\cite{Chang2014, Wang2025ScalablePhotonicQT, OBrien2009PhotonicQuantumTechnologies} necessitates the efficient, scalable integration of identical quantum emitters. Neutral atoms offer a compelling alternative to solid-state emitters for PIC-based interfaces, providing exceptionally long coherence times, perfect indistinguishability, and pristine controllability. Coupling neutral atoms to PIC architectures not only enriches hybrid quantum system functionalities toward deterministic gate-based operations~\cite{1997PRL_transfer, Duan2004ScalablePhotonicQC, Kimble2008QuantumInternet, Chen2013AllOpticalTransistor, Reiserer2014AtomPhotonGate,2014Science_photonrouting, Bechler2018PhotonAtomSwap} but also unlocks new pathways for exploring quantum many-body physics with strongly interacting atoms and photons~\cite{chang2018colloquium, 2023rmp_wQED, GonzalezTudela2024QuantumNanophotonics, Lodahl2017ChiralQuantumOptics,hansen2026realization-142}. In turn, merging cold-atom systems with PIC technologies~\cite{corsetti2026integrated,christen2025integrated,blumenthal2024enabling} leverages established CMOS manufacturing infrastructures, offering a viable road map toward scaling and miniaturization of atom-based quantum hardware.

Interfacing cold atoms with PICs requires efficient loading and stable trapping mechanisms in the immediate vicinity of on-chip photonic structures. While standard magneto-optical trapping (MOT) techniques have been utilized to collect cold atoms around suspended photonic waveguides~\cite{vetsch2010optical,goban2012demonstration, Goban_NatCommun_2014, Goban_PRL_2015, 2016PRL_Jappel, 2016PRL_laurat} and cavities~\cite{2013Science_singleatom,2014Nature_phaseswitch,2021science_entanglementtransport}, achieving MOT directly on a PIC remains elusive due to the restricted optical access and surface reflections (and scattering) inherent to planar circuit geometries. To circumvent this constraint, the current experimental schemes rely on intermittent loading protocols to deliver cold atoms into a tight optical trap on a circuit. For example, tightly focused optical beams have guided both single atoms~\cite{Zhou2023CouplingGuiding} and atomic ensembles~\cite{2024PRX_trappedatoms} into the evanescent field region of microring resonators for collective atom-photon coupling~\cite{2024PRX_trappedatoms, zhou2025selective}. Similarly, arrays of optical tweezers have steered cold atoms onto suspended photonic structures for background-free imaging~\cite{Menon2024} or detecting atom-surface interactions~\cite{hansen2025deterministic}. Recently, cold atoms have been launched from an atomic fountain, followed by probabilistic single-photon cooling and capture within an optical tweezer focused on a microring, realizing strong atom-photon coupling~\cite{margalit2026single}.

Optical conveyor belts~\cite{kuhr2001deterministic,2003PRL_CoherenceProperties,2005PRL_singleatomintocavity} offer an alternative paradigm for the \emph{continuous} delivery of cold atoms, from a separate MOT to areas for strong atom-photon coupling, by using a one-dimensional moving optical lattice. For instance, a conveyor belt can transport atoms toward a suspended photonic crystal waveguide to interrogate atom-light interactions near a photonic bandgap~\cite{2019PNAS_Kimble}. Similarly, translating optical tweezers provide localized delivery and subsequent fluorescence imaging of trapped atoms within a few micrometers above a PIC surface~\cite{2019NC_MayKim,2025PRApplied_Leixu}. However, whether these techniques can be adapted to deliver single atoms and achieve stable trapping for deterministic strong coupling has remained an open question.

\begin{figure*}[t]
\centering
\includegraphics[width=1.5\columnwidth]{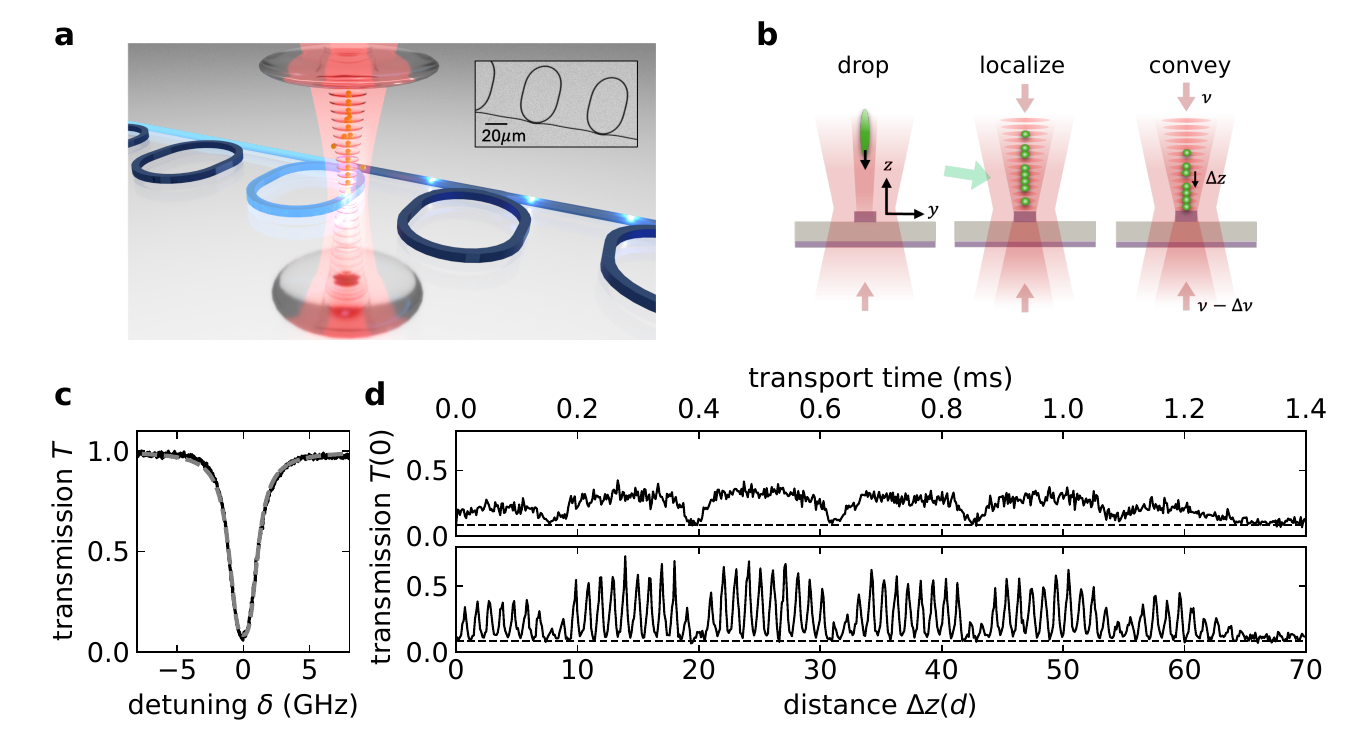}
\caption{\textbf{Conveyor-belt transport from a cold atom reservoir.} 
\textbf{a}, Schematic of the experimental setup. An optical conveyor belt, formed by a moving optical lattice in an optical tweezer, transports a steady stream of trapped atoms onto a microring resonator. The microring draws in resonant photons from a bus waveguide. Atom-induced transparency in cavity QED generates light pulses synchronized with the conveyor-belt movement. Inset shows a scanning electron micrograph of the microring array. \textbf{b}, Experimental procedure depicting the initial atom guiding (drop), loading into an optical lattice (localize) with additional cooling (green arrow), and the subsequent conveyor-belt transport (convey) controlled by the detuning $\Delta \nu$ between the two interfering laser beams (top and bottom red arrows). \textbf{c}, Measured (black line) and fitted (gray dashed line) transmission spectra of the bare microring. \textbf{d}, Measured resonant transmission $T(\delta =0)$ during the conveyor-belt transport. The signal is averaged over 180 repetitions with random starting positions (top). After correcting for the initial position drifts, the data shows fully synchronized, site-resolved transparency pulses (bottom). Lattice spacing $d=467~$nm.}
\label{fig:fig1}
\end{figure*}

Here, we report the first demonstration of near-deterministic single-atom loading on a microring PIC, reaching strong-coupling regime in cavity quantum electrodynamics (QED) with single-atom cooperativity $C > 1$. We realize an optical conveyor belt formed inside an optical tweezer, delivering cesium atoms on-demand from a reservoir to a microring that has been thermally tuned to resonantly couple with the cesium D2 line at $\lambda\approx 852~$nm. Transported atoms at distance $z\ll \lambda$ above the microring evanescently couple to the resonator mode, inducing strong transparency of photons and allowing us to resolve the position of approaching lattice sites down to $4~$nm precision. We further demonstrate near-deterministic single-atom loading by transferring detected atoms into a stationary trap located at $170~$nm above the microring. We report $\gtrsim97\%$ success rate for trap transfer, with a single-atom (two-atom) loading probability $P_1\approx 82(5)\%$ ($P_2\approx 18(6)\%$). The trapped atoms can be continuously probed with an estimated single atom-photon coupling rate $g \approx 2\pi\times 48(1)~$MHz and cooperativity $C=4g^2/(\kappa\Gamma)\approx 1.04(4)$, where $\Gamma=2\pi\times 5.22~$MHz is the single atom decay rate in free space and $\kappa\approx 2\pi\times  1.71~$GHz the resonator photon loss rate. 

\section{Loading and probing trapped atoms on a photonic integrated circuit}
The experimental setup is illustrated in Fig.~\ref{fig:fig1}(a-b). The PIC comprises an array of top-vacuum-cladded silicon-nitride microring resonators fabricated on a silica substrate. Each microring is near-critically coupled to a bus waveguide, which is edge-coupled at both ends to a pair of cleaved optical fibers routed through the vacuum chamber~\cite{Zhou2023CouplingGuiding}. An optical tweezer of wavelength $\lambda_\mathrm{r}\approx 935~$nm is focused to a $2~\mu$m spot on a microring whose waveguide width is $950~$nm. A bottom-illuminating beam of $10~\mu$m waist, derived from the same laser, transmits through the transparent substrate and interferes with the optical tweezer with a beat-note frequency $\Delta \nu$ controlled by an arbitrary waveform generator; $\Delta\nu$ is initially set to zero. The laser beams are far red-detuned from the atomic transition. As a result, cold atoms are attracted to the antinodes of the interference fringes, which form an optical lattice. Another laser excites the microring at wavelength $\lambda_\mathrm{b}\approx 850~$nm, which is one free-spectral-range away and blue-detuned from the atomic resonance. Its evanescent field forms a steep repulsive potential, preventing atoms from crashing onto the microring due to the Casimir-Polder attraction~\cite{Zhou2023CouplingGuiding,2024PRX_trappedatoms}.

\begin{figure}[t]
\centering
\includegraphics[width=1.0\columnwidth]{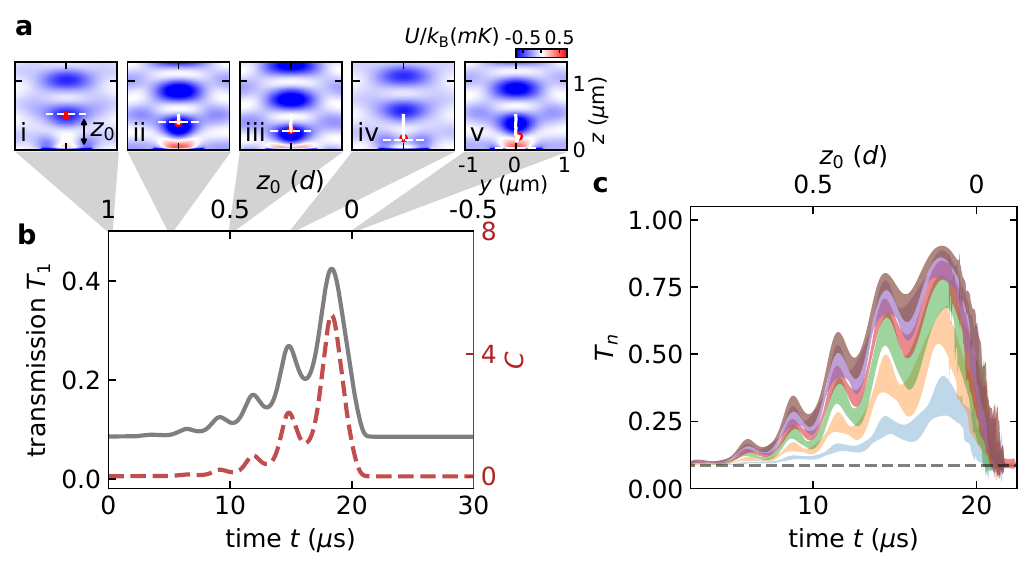}
\caption{\textbf{Simulated atom-photon coupling during transport.}
\textbf{a}, Potential cross-sections $U(y,z)$ overlaid with a representative atomic trajectory (white lines, plotted from the initial time to the indicated time), where $k_\mathrm{B}$ is the Boltzmann constant. The marked atomic positions (red circles) follow the occupied lattice site (marked by dashed lines), as the distance of the site center to the surface ($z_0$) moves from $z_0/d=$1 (i) to 0.75 (ii), 0.5 (iii), 0.25 (iv), and 0 (v). \textbf{b}, Calculated resonant transmission $T_1$ (gray line) and single-atom cooperativity $C$ (red dashed line) during the transport, both averaged over 10 single-atom trajectories. \textbf{c}, Transmission $T_n$ for $n=1,2,\ldots, 6$ atoms confined in a single site (colored curves, from lowest to highest). Shaded bands indicate the standard deviation calculated from simulated trajectories. 
}
\label{fig:fig2}
\end{figure}

\begin{figure}[t]
\centering
\includegraphics[width=1.0\columnwidth]{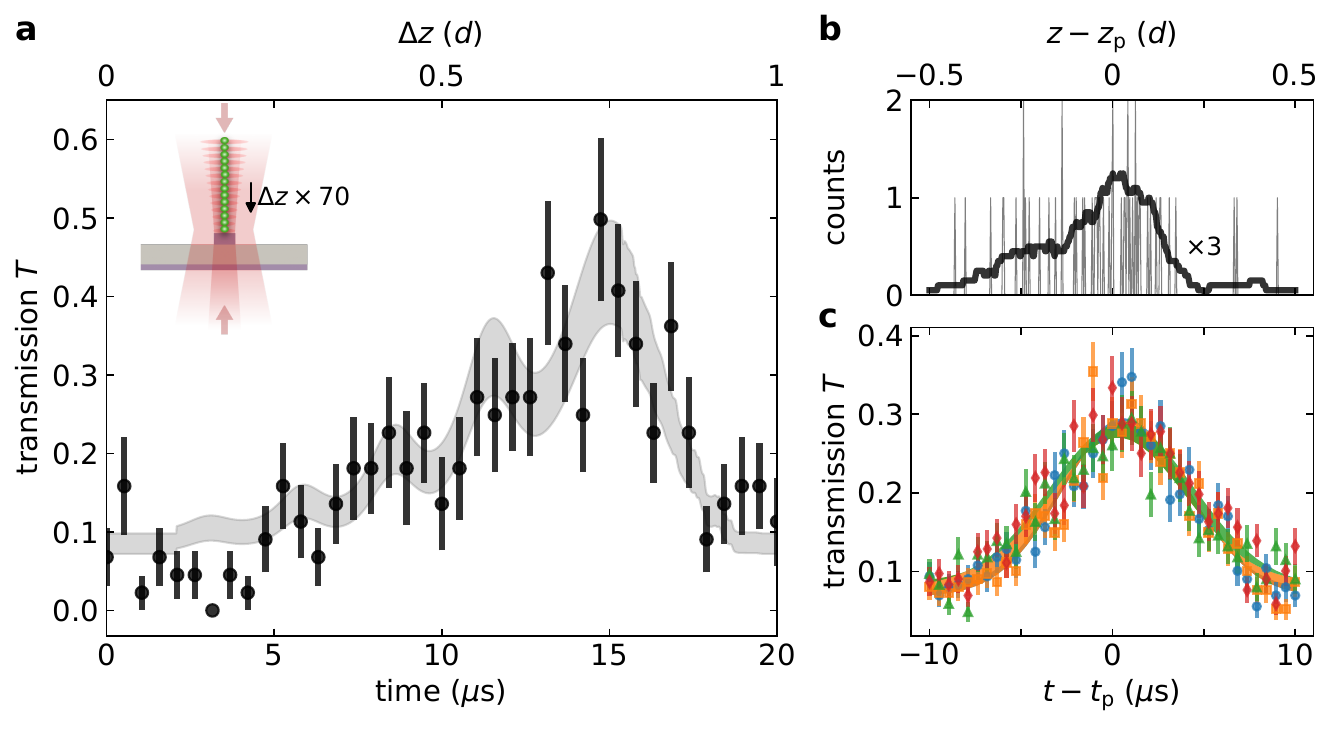}
\caption{\textbf{Site-resolved measurement.}
\textbf{a}, Averaged transmission $T$ of 70 lattice sites in one transport. Shaded band shows the fitted transmission curve and its uncertainty, based on the model $T_n$ in Fig.~\ref{fig:fig2}\textbf{c} and weighted averaged using Poisson atom-number statistics. The fit yields a mean site occupancy $\bar{n}=1.5(5)$. \textbf{b}, Time-resolved detector counts (per 10~ns bin) accumulated while transporting the first 10 lattice sites. The transmission peak at $t=t_\mathrm{p}$ ($z=z_\mathrm{p}$) is clearly visible in the average over a 0.6~$\mu$s running window (solid line). \textbf{c}, Averaged transmission of four conveyor-belt transports (color symbols), demonstrating excellent alignment using the 10-site average as shown in \textbf{b} to determine the random offset $t_\mathrm{p}$ ($z_\mathrm{p}$). Color lines show Gaussian fits to each measurement, giving the standard deviation of the peak positions $\sigma_z  =0.008d$. Error bars in \textbf{a} and \textbf{c} denote the standard error of the mean.
}
\label{fig:fig3}
\end{figure}

We begin the experiment by releasing laser-cooled atoms, initially positioned at $\sim100~\mu$m above the PIC, and guiding them onto the microring using the bottom beam~\cite{Zhou2023CouplingGuiding}. After a 5~ms fall toward the surface, we switch on the optical tweezer to form a static lattice potential. We employ a second cooling stage, called degenerate Raman sideband cooling (dRSC)~\cite{Vuletic1998DegenerateDensities}, to remove the energy acquired during the drop (Methods). After 10~ms of dRSC, atoms fully localize in the lattice with an estimated temperature $\lesssim 20~\mu$K. We then linearly ramp on $\Delta \nu$ in 1~ms to initiate the conveyor-belt transport. Upon reaching a static detuning $\Delta \nu=50~$kHz, the lattice sites move toward the PIC surface at a constant speed of $d\Delta \nu \approx 23$~nm/$\mu$s, where $d=\lambda_\mathrm{r}/2 \approx 467~$nm is the lattice spacing.

We record atom-induced transparency versus transport time $t$ (and moving distance $\Delta z=d\Delta \nu t$) 
by monitoring the transmission of a weak probe through the bus waveguide. In the absence of atoms, the resonant transmission $T$ at zero detuning is nearly diminished, as shown in Fig.~\ref{fig:fig1}(c), contrasting the case with trapped atoms loaded in the conveyor belt. The top panel of Fig.~\ref{fig:fig1}(d) shows an average of transmission $T$ from repeated experiments, revealing the presence of trapped atoms in roughly $70$ lattice sites along the conveyor belt. We observe notches in every 11 sites, indicating regions of low loading efficiency. These regions coincide with the interference nodes in the dRSC cooling beam, formed as the beam illuminates the PIC at a 5$^\circ$ grazing angle and interferes with its surface reflection.

Ideally, the atom-induced transparency should reveal the lattice structure in the conveyor belt. This is nevertheless averaged out due to long-term phase fluctuations in the laser beams, which randomize the starting lattice position in each transport. We correct for the drifts by directly using few-site transparency signal; see also Fig.~\ref{fig:fig3} and discussions. In the bottom panel of Fig.~\ref{fig:fig1}(d), we plot the realigned signal from the same set of measurements, revealing sharp transparency `pulses' due to the expected lattice movement. The majority of sites exhibit peak $T>0.5$, hinting that trapped atoms can approach strong coupling near the microring surface.

\begin{figure*}[t]
\centering
\includegraphics[width=1.5\columnwidth]{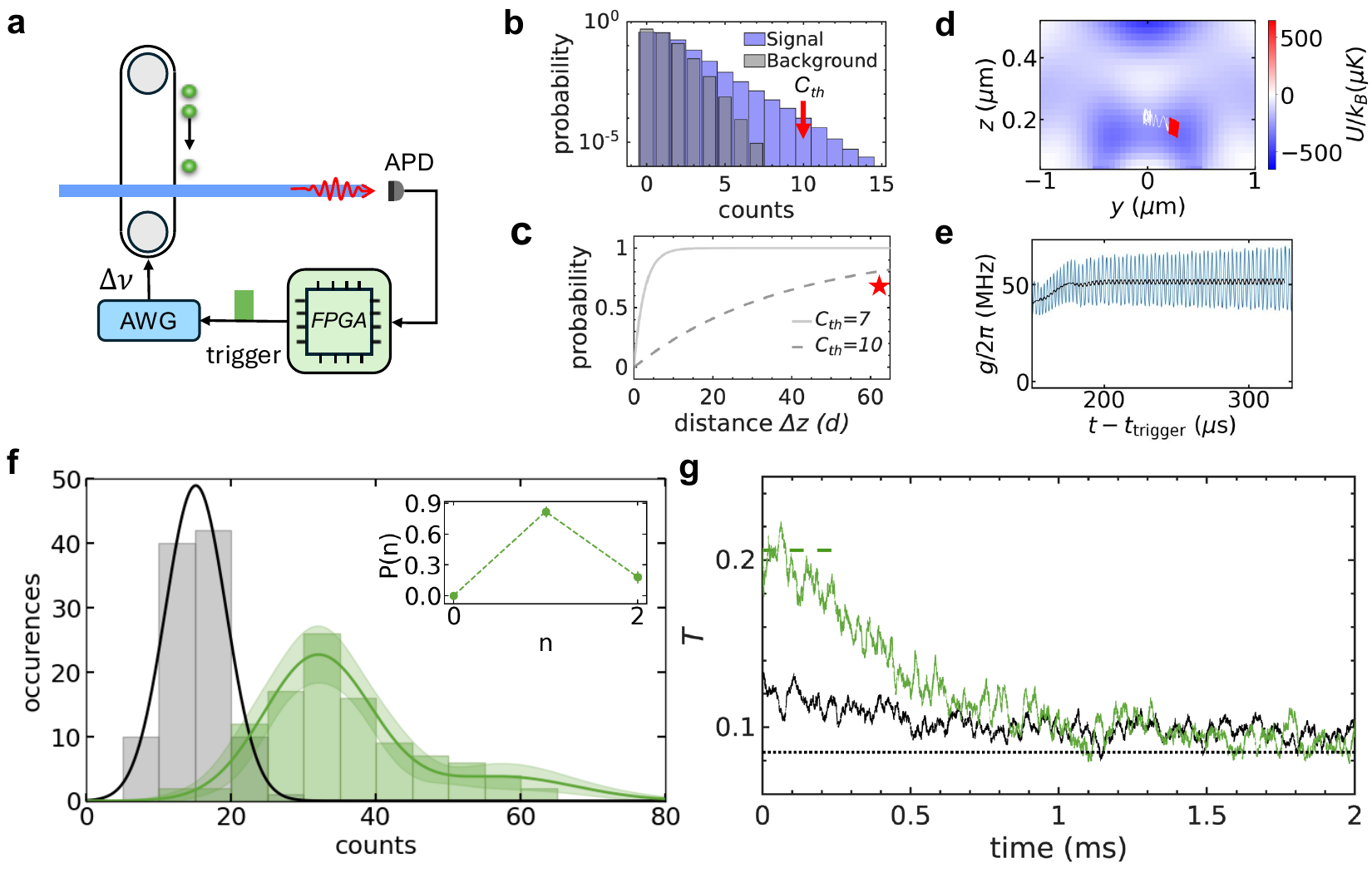}
\caption{\textbf{Near-deterministic single-atom loading.}
\textbf{a}, Schematic of feedback-controlled conveyor-belt transport. An FPGA monitors the time-averaged counts from the avalanche photodiode (APD) and triggers the arbitrary waveform generator (AWG) to halt the transport by ramping down $\Delta\nu$ while ramping up the tweezer power. \textbf{b}, Probability histogram of the APD counts in 10~$\mu$s bins. Trigger threshold $C_\mathrm{th}$ is set to values with diminishing probability from background counts. \textbf{c}, Calculated (gray lines) and measured (red star) trigger probability versus transport distance $\Delta z$. \textbf{d}, The final stationary potential cross-section and a simulated trajectory plotted following the trigger (white line) and after the trap becomes fully stationary (red line). \textbf{e}, The corresponding atom-photon coupling strength $g$ (blue line) for the trajectory in \textbf{d}. Black line is a 10~$\mu$s running average. \textbf{f}, Histogram of APD counts collected in a $200~\mu$s time period after the trap transfer (green bars) and the background counts without atoms (gray bars). Solid lines are fits (see text). Inset shows the probability $P_n$ of loading $n=0,1,2$ atoms. Error bars show the fit uncertainty. \textbf{g}, Continuously probing the loaded atoms following feedback control (green curve) or with random stops (black curve). Green dashed line marks the averaged atom-induced transmission calculated from the fit in \textbf{f}. Black dotted line denotes the background.
}
\label{fig:fig4}
\end{figure*}

We analyze how the pulses reveal strong atom-photon coupling through modeling single-atom trap dynamics in the conveyor belt. Figure~\ref{fig:fig2}(a) displays snapshots of a simulated atom trajectory, as the lattice potential sweeps through the evanescent field region until the trap deforms and allows the atom to escape. From 10 randomly sampled single-atom trajectories, we calculate the averaged transmission $T_1$ as shown in Fig.~\ref{fig:fig2}(b), which rises above $0.25$ when the distance of the site center to the surface ($z_0$) moves toward $d/4 \approx 117~$nm, below which point the trap opens. The transmission peaks above 0.4 and quickly drops to the background level $T_0\approx 0.085$ after the atom is lost. During the transit, the cooperativity $C$ reaches beyond $4$. Visible oscillations in $C$ ($T_1$) are caused by the quasiperiodic trap motion along the $z$-axis (frequency $\sim 300~$kHz), which should average out in a large ensemble of random trajectories.

Our measured peaks $T>0.5$ suggest the possibility of reaching site occupancy $n>1$. In Fig. \ref{fig:fig2}(c), we similarly calculate the transmission $T_n$ of $n$ atoms simultaneously loaded into a single lattice site. Overall, the pulse shapes remain qualitatively similar, but the $n$-atom cooperative coupling leads to resolvable, higher transparency with increasing atom number. Comparing the magnitude of the peaks with the measured signal in Fig.~\ref{fig:fig1}(d), the actual site occupancy likely ranges between $n\approx 1\sim 2$ where peak $T_n\approx 0.4\sim 0.7$.

The synchronization of pulses observed in Fig.~\ref{fig:fig1}(d) allows us to resolve single-site dynamics and extract the mean site occupancy. In Fig.~\ref{fig:fig3}(a), we plot the transmission pulse averaged over 70 sites from one transport. The data is plotted (and shifted) within one period of $1/\Delta \nu=20~\mu$s. Remarkably, the signal shows excellent agreement with the simulated line shape. We can determine the mean site occupancy $\bar{n} = 1.5(5)$ by fitting the signal with a Poisson-weighted sum of $T_n$ curves as shown in Fig.~\ref{fig:fig2}(c). This suggests the majority of delivered sites contain one to two atoms reaching strong coupling with the resonator and only 20\% of the sites are empty. 

To further verify that the site occupancy is indeed close to unity, we have performed resonance fluorescence and photon correlation measurements. Our observation of non-classical photon correlation function at zero delay time, $g^{(2)}(0)<0.5$, confirms that the conveyor belt is primarily loaded with single atoms; See Methods for detailed discussions on the $g^{(2)}$ measurement.

\section{Precision conveyor-belt transport and near-deterministic single-atom loading}
It is desirable to control the exact timing and position of arriving lattice sites. Achieving this goal with long-term stability, however, requires either actively stabilizing the optical phases or tagging the lattice position through scattered light in the circuit~\cite{2019PNAS_Kimble}. Here, we demonstrate a simple and effective scheme, by showing that few-site transmission signal is sufficient to pinpoint the lattice positions accurately. In Fig.~\ref{fig:fig3}(b), we plot the time trace of single-photon detector counts and its running average accumulated over the first 10 lattice sites. A pulse structure clearly emerges, similar to the line shape shown in Fig.~\ref{fig:fig3}(a). Through the running average, we can identify the timing $t_\mathrm{p}$ and position $z_\mathrm{p}$ of the peak. Thus, the arrival time (distance) of all remaining sites can be determined, $t_j = t_\mathrm{p} + j/\Delta \nu$ ($z_j = z_\mathrm{p} + jd$), where $j$ is the site index. Figure~\ref{fig:fig3}(c) shows four independent transports, where we identify the respective $t_\mathrm{p}$ ($z_\mathrm{p}$) and plot the average from the remaining 60 sites in each transport. There appears to be very small error in the detected peak positions. We estimate the error (standard deviation) of the peaks to be $\sigma_z =0.008d\approx 4~$nm, corresponding to a timing error of $\sigma_t\approx 160~$ns. 

We note that our demonstrated few-site peak detection protocol can be implemented in real time using a field-programmable gate array (FPGA), thus enabling a self-aligned conveyor belt transport. The expected hardware delay from the peak search is smaller than the timing uncertainty $\sigma_t$ and can be offset deterministically in the program. 

In addition to the clocked delivery scheme, we also utilize active feedback to achieve near-deterministic single-atom trapping on a microring. To date, feedback trapping in the evanescent field region has remained technically challenging. This is because, for free atoms, the interaction time is typically limited to just a few microseconds \cite{Aoki2006,2014Science_photonrouting,2021PRL_TrapWGM, Zhou2023CouplingGuiding}. Employing an optical conveyor belt, on the other hand, this timing can be very well controlled. As illustrated in Fig.~\ref{fig:fig4}(a), we start by a smaller $\Delta\nu=6$ kHz for slow conveyor-belt motion $d\Delta\nu \approx 2.8~$nm/$\mu$s, extending the interaction time to tens of microseconds. We deploy an FPGA to monitor the transmission counts in real time. Upon exceeding a threshold, the FPGA triggers ramp-down of the detuning $\Delta \nu$ (typically within $50~\mu$s) to smoothly halt the conveyor belt without heating out the trapped atoms. We choose a proper threshold by analyzing the histogram of the counts (per $10~\mu$s bin) prerecorded in a continuous delivery, as shown in Fig.~\ref{fig:fig4}(b), and set the threshold to $C_\mathrm{th}=10$ to ensure high-fidelity atom trapping and sufficient feedback probability within our conveyor-belt moving range. In the experiment, we measured the probability for triggering to be $69~$\% after moving over 60 populated sites, in good agreement with the expectation shown in Fig.~\ref{fig:fig4}(c).  

Upon triggering, we transfer the detected atoms into a stationary trap located at $(y, z)\approx (\pm 245, 170)~$nm, as shown in Fig.\ref{fig:fig4}(d), by ramping up the tweezer power within $5~\mu$s to immediately override the conveyor-belt potential. The trap is formed by a stable interference pattern between the tweezer and its surface reflection (see Methods). We have simulated this process in Monte Carlo simulations, rendering nearly no failure in trap transfer using our experimental parameters. Figure~\ref{fig:fig4}(d) shows a sample trajectory, plotted since the trigger till after the trap transfer. Figure~\ref{fig:fig4}(e) shows the calculated atom-photon coupling strength during this time, giving a mean $g \approx2\pi\times51\,\mathrm{MHz}$ and $C\approx 1.2$.

We experimentally confirm successful atom trapping by continuously probing the transmission after completion of the trap transfer. In Fig.~\ref{fig:fig4}(f), the histogram of counts detected in $200\,\mu\mathrm{s}$ reveals contribution primarily from single trapped atoms. The measured peak location relative to the background in the histogram corresponds to $T\approx 0.19$, in perfect agreement with the expectation from single trapped atoms. Using a composite Gaussian fit with input from the expected mean $\bar{T}_n\approx 0.085, 0.20, 0.35$ for $n=0, 1, 2$ trapped atoms (Methods), we find the single-(two-)atom trapping probability $P_1\approx 0.82(5)$ ($P_2 \approx 0.18(6)$) and mean occupancy $\bar{n}\approx 1.2$. From the fit, we determine a slightly reduced $g\approx2\pi \times 48(1)~$MHz and $C\approx 1.04(4)$. Remarkably, we recorded $\gtrsim 97$\% success rate for the trap transfer, conditioned upon receiving a trigger. The success is dominated by single-atom trapping. The reduced two-atom probability is likely due to inelastic collision loss during the trap transfer and probe procedures.

We further exemplify the effectiveness of feedback control in Fig.~\ref{fig:fig4}(g), where we compare the transmission signal with those obtained from randomly stopping the lattice potential. In the latter case, random and unstable final trap condition near the surface renders a significantly lower transparency signal. On the other hand, the stably transferred atoms present $T\approx \bar{T}_1\approx 0.2$ (for $C\approx 1$) and can be continuously probed with a $1/e$ lifetime around 500~$\mu$s, which is expected from resonant heating \cite{2026Aoki_array,2026nearfieldheating,margalit2026single}. Lastly, we note that the measured lifetime of trapped atoms near the surface (without continuous probing and resonant heating) is over 160~ms, similar to that of the trapped atomic ensemble experiment reported in Ref. \cite{2024PRX_trappedatoms}; See Methods for details. 

\section{Discussion and outlook}
To summarize, we present the first conveyor-belt delivery of single atoms with nanoscale precision onto a functional PIC, realizing strong atom-photon coupling $C\gtrsim1$ for cavity QED on a PIC. Utilizing an FPGA and simple feedback control, we demonstrate deterministic single- to two-atom trapping in a stable trap $\approx 170$~nm above the surface of a microring. In addition, the demonstrated conveyor belt transport can self-align with a 4~nm precision. Our demonstration of precision single-atom delivery from an atomic reservoir thus uniquely points to a possible pathway for deterministic and even continuous atom-array assembly using a conveyor-belt enabled optical tweezer array. Realizing arrays of trapped atoms strongly coupled to a PIC promises future applications in quantum communications, quantum simulations, and even photonic quantum computing using programmable integrated circuits.


\bibliography{apssamp}

@article{2003PRL_CoherenceProperties,
  title = {Coherence Properties and Quantum State Transportation in an Optical Conveyor Belt},
  author = {Kuhr, S. and Alt, W. and Schrader, D. and Dotsenko, I. and Miroshnychenko, Y. and Rosenfeld, W. and Khudaverdyan, M. and Gomer, V. and Rauschenbeutel, A. and Meschede, D.},
  journal = {Phys. Rev. Lett.},
  volume = {91},
  issue = {21},
  pages = {213002},
  numpages = {4},
  year = {2003},
  month = {Nov},
  publisher = {American Physical Society},
  doi = {10.1103/PhysRevLett.91.213002},
  url = {https://link.aps.org/doi/10.1103/PhysRevLett.91.213002}
}

@article{
2019PNAS_Kimble,
author = {A. P. Burgers  and L. S. Peng  and J. A. Muniz  and A. C. McClung  and M. J. Martin  and H. J. Kimble },
title = {Clocked atom delivery to a photonic crystal waveguide},
journal = {Proceedings of the National Academy of Sciences},
volume = {116},
number = {2},
pages = {456-465},
year = {2019},
doi = {10.1073/pnas.1817249115},
URL = {https://www.pnas.org/doi/abs/10.1073/pnas.1817249115}}

@Article{2019NC_MayKim,
author={Kim, May E.
and Chang, Tzu-Han
and Fields, Brian M.
and Chen, Cheng-An
and Hung, Chen-Lung},
title={Trapping single atoms on a nanophotonic circuit with configurable tweezer lattices},
journal={Nature Communications},
year={2019},
month={Apr},
day={09},
volume={10},
number={1},
pages={1647},
issn={2041-1723},
doi={10.1038/s41467-019-09635-7},
url={https://doi.org/10.1038/s41467-019-09635-7}
}

@article{2005PRL_singleatomintocavity,
  title = {Submicron Positioning of Single Atoms in a Microcavity},
  author = {Nu\ss{}mann, Stefan and Hijlkema, Markus and Weber, Bernhard and Rohde, Felix and Rempe, Gerhard and Kuhn, Axel},
  journal = {Phys. Rev. Lett.},
  volume = {95},
  issue = {17},
  pages = {173602},
  numpages = {4},
  year = {2005},
  month = {Oct},
  publisher = {American Physical Society},
  doi = {10.1103/PhysRevLett.95.173602},
  url = {https://link.aps.org/doi/10.1103/PhysRevLett.95.173602}
}

@article{Zhou2023CouplingGuiding,
  title = {Coupling Single Atoms to a Nanophotonic Whispering-Gallery-Mode Resonator via Optical Guiding},
  author = {Zhou, Xinchao and Tamura, Hikaru and Chang, Tzu-Han and Hung, Chen-Lung},
  journal = {Phys. Rev. Lett.},
  volume = {130},
  issue = {10},
  pages = {103601},
  numpages = {7},
  year = {2023},
  month = {Mar},
  publisher = {American Physical Society},
  doi = {10.1103/PhysRevLett.130.103601},
  url = {https://link.aps.org/doi/10.1103/PhysRevLett.130.103601}
}

@article{Vuletic1998DegenerateDensities,
  title = {Degenerate Raman Sideband Cooling of Trapped Cesium Atoms at Very High Atomic Densities},
  author = {Vuleti\ifmmode \acute{c}\else \'{c}\fi{}, Vladan and Chin, Cheng and Kerman, Andrew J. and Chu, Steven},
  journal = {Phys. Rev. Lett.},
  volume = {81},
  issue = {26},
  pages = {5768--5771},
  numpages = {0},
  year = {1998},
  month = {Dec},
  publisher = {American Physical Society},
  doi = {10.1103/PhysRevLett.81.5768},
  url = {https://link.aps.org/doi/10.1103/PhysRevLett.81.5768}
}

@article{2021PRL_TrapWGM,
  title = {Coupling a Single Trapped Atom to a Whispering-Gallery-Mode Microresonator},
  author = {Will, Elisa and Masters, Luke and Rauschenbeutel, Arno and Scheucher, Michael and Volz, J\"urgen},
  journal = {Phys. Rev. Lett.},
  volume = {126},
  issue = {23},
  pages = {233602},
  numpages = {5},
  year = {2021},
  month = {Jun},
  publisher = {American Physical Society},
  doi = {10.1103/PhysRevLett.126.233602},
  url = {https://link.aps.org/doi/10.1103/PhysRevLett.126.233602}
}

@article{2014Science_photonrouting,
author = {Itay Shomroni  and Serge Rosenblum  and Yulia Lovsky  and Orel Bechler  and Gabriel Guendelman  and Barak Dayan },
title = {All-optical routing of single photons by a one-atom switch controlled by a single photon},
journal = {Science},
volume = {345},
number = {6199},
pages = {903-906},
year = {2014},
doi = {10.1126/science.1254699},
URL = {https://www.science.org/doi/abs/10.1126/science.1254699},
}

@Article{Aoki2006,
author={Aoki, Takao
and Dayan, Barak
and Wilcut, E.
and Bowen, W. P.
and Parkins, A. S.
and Kippenberg, T. J.
and Vahala, K. J.
and Kimble, H. J.},
title={Observation of strong coupling between one atom and a monolithic microresonator},
journal={Nature},
year={2006},
month={Oct},
day={01},
volume={443},
number={7112},
pages={671-674},
issn={1476-4687},
doi={10.1038/nature05147},
url={https://doi.org/10.1038/nature05147}
}

@article{2024PRX_trappedatoms,
  title = {Trapped Atoms and Superradiance on an Integrated Nanophotonic Microring Circuit},
  author = {Zhou, Xinchao and Tamura, Hikaru and Chang, Tzu-Han and Hung, Chen-Lung},
  journal = {Phys. Rev. X},
  volume = {14},
  issue = {3},
  pages = {031004},
  numpages = {11},
  year = {2024},
  month = {Jul},
  publisher = {American Physical Society},
  doi = {10.1103/PhysRevX.14.031004},
  url = {https://link.aps.org/doi/10.1103/PhysRevX.14.031004}
}

@article{Menon2024,
    title = {An integrated atom array-nanophotonic chip platform with background-free imaging},
    author = {Menon, Shankar G. and Glachman, Noah and Pompili, Matteo and Dibos, Alan and Bernien, Hannes},
    journal = {Nature Communications},
    volume = {15},
    issue = {1},
    pages = {6156},
    year = {2024},
    month = {Jul},
    doi = {10.1038/s41467-024-50355-4},
    url = {https://doi.org/10.1038/s41467-024-50355-4},    
}

@article{Goban_NatCommun_2014,
  title = {Atom–light interactions in photonic crystals},
  author = {Goban, A. and Hung, C.-L. and Yu, S.-P. and Hood, J.D. and Muniz, J.A. and Lee, J.H. and Martin, M.J. and McClung, A.C. and Choi, K.S. and Chang, D.E. and Painter, O. and Kimble, H.J.},
  journal = {Nature Communications},
  volume = {5},
  issue = {1},
  pages = {3808},
  numpages = {},
  year = {2014},
  month = {May},
  publisher = {},
  doi = {10.1038/ncomms4808},
  url = {https://doi.org/10.1038/ncomms4808}
}

@article{Goban_PRL_2015,
  title = {Superradiance for Atoms Trapped along a Photonic Crystal Waveguide},
  author = {Goban, A. and Hung, C.-L. and Hood, J. D. and Yu, S.-P. and Muniz, J. A. and Painter, O. and Kimble, H. J.},
  journal = {Phys. Rev. Lett.},
  volume = {115},
  issue = {6},
  pages = {063601},
  numpages = {5},
  year = {2015},
  month = {Aug},
  publisher = {American Physical Society},
  doi = {10.1103/PhysRevLett.115.063601},
  url = {https://link.aps.org/doi/10.1103/PhysRevLett.115.063601}
}

@article{chang2018colloquium,
  title = {Colloquium: Quantum matter built from nanoscopic lattices of atoms and photons},
  author = {Chang, D. E. and Douglas, J. S. and Gonz\'alez-Tudela, A. and Hung, C.-L. and Kimble, H. J.},
  journal = {Rev. Mod. Phys.},
  volume = {90},
  issue = {3},
  pages = {031002},
  numpages = {30},
  year = {2018},
  month = {Aug},
  publisher = {American Physical Society},
  doi = {10.1103/RevModPhys.90.031002},
  url = {https://link.aps.org/doi/10.1103/RevModPhys.90.031002}
}

@article{hansen2025deterministic,
  title={Deterministic coupling of ultracold atomic lattice to a suspended photonic waveguide},
  author={Hansen, JT and Gargiulo, F and Mathiassen, JB and M{\"u}ller, JH and Polzik, ES and B{\'e}guin, J-B},
  journal={arXiv:2511.18211},
  year={2025},
  url={https://arxiv.org/abs/2511.18211}
}

@article{zhou2025selective,
    title = {Selective Collective Emission from a Dense Atomic Ensemble Coupled to a Nanophotonic Resonator},
  author = {Zhou, Xinchao and Suresh, Deepak A. and Robicheaux, F. and Hung, Chen-Lung},
  journal = {Phys. Rev. Lett.},
  volume = {135},
  issue = {11},
  pages = {113601},
  numpages = {7},
  year = {2025},
  month = {Sep},
  publisher = {American Physical Society},
  doi = {10.1103/cdd5-r7h4},
  url = {https://link.aps.org/doi/10.1103/cdd5-r7h4}
}

@article{margalit2026single,
  title={Single-atom trapping in the evanescent field of an integrated photonic resonator}, 
  author={Yair Margalit and Omri Davidson and Oded Zemer and Yoad Michael and Orel Bechler and Dror Liran and Noam Gross and Doron Azoury and Jeremy Raskop and Yaakov Yudkin and Gabriel Guendelman and Moshe Katzman and Michael Nagli and Yair Antman and Nadav Kandel and Geva Arwas and Idit Peer and Ofer Firstenberg and Barak Dayan},
  year={2026},
  journal={arXiv:2605.09532},
  archivePrefix={arXiv},
  primaryClass={quant-ph},
  url={https://arxiv.org/abs/2605.09532},   
}

@article{kuhr2001deterministic,
  author = {Kuhr, Stephan and Alt, Wolfgang and Schrader, Dominik and M{\"u}ller, Martin and Gomer, Volker and Meschede, Dieter},
  title = {Deterministic Delivery of a Single Atom},
  journal = {Science},
  volume = {293},
  number = {5528},
  pages = {278--280},
  year = {2001},
  doi = {10.1126/science.1062725},
  publisher = {American Association for the Advancement of Science (AAAS)}
}

@article{hung2013trapped,
  author = {Hung, Chen-Lung and Meenehan, Se{\'a}n M. and Chang, Darrick E. and Painter, Oskar and Kimble, H. Jeff},
  title = {Trapped atoms in one-dimensional photonic crystals},
  journal = {New Journal of Physics},
  volume = {15},
  number = {8},
  pages = {083026},
  year = {2013},
  doi = {10.1088/1367-2630/15/8/083026},
  publisher = {IOP Publishing}
}

@article{2026Aoki_array,
    title={Fiber-optic quantum interface with an array of more than 100 individually addressable atoms on an optical nanofiber}, 
    author={Mitsuyoshi Takahata and Jameesh Keloth and Takashi Yamamoto and Ken-ichi Harada and Shigehito Miki and Takao Aoki},
    journal={arXiv: 2603.21812},
    year={2026},
    archivePrefix={arXiv},
    url={https://arxiv.org/abs/2603.21812}, 
}

@article{Chang2014,
    author = {Chang, Darrick E. and Vuleti{\'c}, Vladan and Lukin, Mikhail D.},
    title = {Quantum nonlinear optics — photon by photon},
    year = {2014},
    month = {Sep},
    journal = {Nature Photonics},
    pages = {685-694},
    volume = {8},
    issue = {9},
    url = {https://doi.org/10.1038/nphoton.2014.192},
    doi = {10.1038/nphoton.2014.192},
}

@article{2023NaturePhotonics_laserchip,
  author  = {Mateus Corato-Zanarella and Andres Gil-Molina and Xingchen Ji and Min Chul Shin and Aseema Mohanty and Michal Lipson},
  title={Widely tunable and narrow-linewidth chip-scale lasers from near-ultraviolet to near-infrared wavelengths},
  author={Corato-Zanarella, Mateus and Gil-Molina, Andres and Ji, Xingchen and Shin, Min Chul and Mohanty, Aseema and Lipson, Michal},
  journal={Nature Photonics},
  volume={17},
  number={2},
  pages={157--164},
  year={2023},
  publisher={Nature Publishing Group UK London},
  doi      = {10.1038/s41566-022-01120-w},
}

@article{2026Nature_ChenFiberLikeLoss,
  author = {Chen, Hao-Jing and Colburn, Kellan and Liu, Peng and Yan, Hongrui and Hou, Hanfei and Ge, Jinhao and Liu, Jin-Yu and Lehan, Phineas and Ji, Qing-Xin and Yuan, Zhiquan and Bouwmeester, Dirk and Holmes, Christopher and Gates, James and Blauvelt, Henry and Vahala, Kerry},
  title = {Towards fibre-like loss for photonic integration from violet to near-infrared},
  journal = {Nature},
  year = {2026},
  volume = {649},
  number = {8096},
  pages = {338--344},
  doi = {10.1038/s41586-025-09889-w},
  publisher = {Springer Nature}
}

@article{2021NaturePhotonics_laser,
  author = {Jin, Warren and Yang, Qi-Fan and Chang, Lin and Shen, Boqiang and Wang, Heming and Leal, Mark A. and Wu, Lue and Gao, Maodong and Feshali, Avi and Paniccia, Mario and Vahala, Kerry J. and Bowers, John E.},
  title = {Hertz-linewidth semiconductor lasers using {CMOS}-ready ultra-high-{Q} microresonators},
  journal = {Nature Photonics},
  volume = {15},
  number = {5},
  pages = {346--353},
  year = {2021},
  doi = {10.1038/s41566-021-00761-7},
  publisher = {Springer Nature}
}

@article{2021science_entanglementtransport,
  author = {Đorđević, Tamara and Samutpraphoot, Polnop and Ocola, Paloma L. and Bernien, Hannes and Grinkemeyer, Brandon and Dimitrova, Ivana and Vuletić, Vladan and Lukin, Mikhail D.},
  title = {Entanglement transport and a nanophotonic interface for atoms in optical tweezers},
  journal = {Science},
  volume = {373},
  number = {6562},
  pages = {1511--1514},
  year = {2021},
  doi = {10.1126/science.abi9917},
  publisher = {American Association for the Advancement of Science (AAAS)}
}

@article{2014Nature_phaseswitch,
  author = {Tiecke, Tobias G. and Thompson, Jeffrey D. and de Leon, Nathalie P. and Liu, Li R. and Vuletić, Vladan and Lukin, Mikhail D.},
  title = {Nanophotonic quantum phase switch with a single atom},
  journal = {Nature},
  volume = {508},
  number = {7495},
  pages = {241--244},
  year = {2014},
  doi = {10.1038/nature13188},
  publisher = {Nature Publishing Group}
}

@article{2013Science_singleatom,
  author = {Thompson, Jeff D. and Tiecke, Tobias G. and de Leon, Nathalie P. and Feist, Johannes and Akimov, Alexey V. and Gullans, Michael and Zibrov, Alexander S. and Vuletić, Vladan and Lukin, Mikhail D.},
  title = {Coupling a Single Trapped Atom to a Nanoscale Optical Cavity},
  journal = {Science},
  volume = {340},
  number = {6137},
  pages = {1202--1205},
  year = {2013},
  doi = {10.1126/science.1237125},
  publisher = {American Association for the Advancement of Science (AAAS)}
}

@article{2024NatPhotonics_review,
  author = {Lu, Xiyuan and Chang, Lin and Tran, Minh A. and Komljenovic, Tin and Bowers, John E. and Srinivasan, Kartik},
  title = {Emerging integrated laser technologies in the visible and short near-infrared regimes},
  journal = {Nature Photonics},
  volume = {18},
  number = {10},
  pages = {1010--1023},
  year = {2024},
  doi = {10.1038/s41566-024-01529-5},
  publisher = {Springer Nature}
}

@article{Lodahl2017ChiralQuantumOptics,
  author = {Lodahl, Peter and Mahmoodian, Sahand and Stobbe, S{\o}ren and Rauschenbeutel, Arno and Schneeweiss, Philipp and Volz, J{\"u}rgen and Pichler, Hannes and Zoller, Peter},
  title = {Chiral quantum optics},
  journal = {Nature},
  volume = {541},
  number = {7638},
  pages = {473--480},
  year = {2017},
  doi = {10.1038/nature21037},
  publisher = {Springer Nature}
}

@article{GonzalezTudela2024QuantumNanophotonics,
author={Gonz{\'a}lez-Tudela, Alejandro
and Reiserer, Andreas
and Garc{\'i}a-Ripoll, Juan Jos{\'e}
and Garc{\'i}a-Vidal, Francisco J.},
title={Light--matter interactions in quantum nanophotonic devices},
journal={Nature Reviews Physics},
year={2024},
month={Mar},
day={01},
volume={6},
number={3},
pages={166-179},
issn={2522-5820},
doi={10.1038/s42254-023-00681-1},
url={https://doi.org/10.1038/s42254-023-00681-1}
}

@article{Kimble2008QuantumInternet,
  author = {Kimble, H. J.},
  title = {The Quantum Internet},
  journal = {Nature},
  volume = {453},
  number = {7198},
  pages = {1023--1030},
  year = {2008},
  doi = {10.1038/nature07127},
  publisher = {Nature Publishing Group}
}

@article{Reiserer2014AtomPhotonGate,
  author = {Reiserer, Andreas and Kalb, Norbert and Rempe, Gerhard and Ritter, Stephan},
  title = {A quantum gate between a flying optical photon and a single trapped atom},
  journal = {Nature},
  volume = {508},
  number = {7495},
  pages = {237--240},
  year = {2014},
  doi = {10.1038/nature13177},
  publisher = {Nature Publishing Group}
}

@article{1997PRL_transfer,
  title = {Quantum State Transfer and Entanglement Distribution among Distant Nodes in a Quantum Network},
  author = {Cirac, J. I. and Zoller, P. and Kimble, H. J. and Mabuchi, H.},
  journal = {Phys. Rev. Lett.},
  volume = {78},
  issue = {16},
  pages = {3221--3224},
  numpages = {0},
  year = {1997},
  month = {Apr},
  publisher = {American Physical Society},
  doi = {10.1103/PhysRevLett.78.3221},
  url = {https://link.aps.org/doi/10.1103/PhysRevLett.78.3221}
}

@article{Duan2004ScalablePhotonicQC,
  title = {Scalable Photonic Quantum Computation through Cavity-Assisted Interactions},
  author = {Duan, L.-M. and Kimble, H. J.},
  journal = {Phys. Rev. Lett.},
  volume = {92},
  issue = {12},
  pages = {127902},
  numpages = {4},
  year = {2004},
  month = {Mar},
  publisher = {American Physical Society},
  doi = {10.1103/PhysRevLett.92.127902},
  url = {https://link.aps.org/doi/10.1103/PhysRevLett.92.127902}
}

@article{Chen2013AllOpticalTransistor,
  author = {Chen, Wenlan and Beck, Kristin M. and B{\"u}cker, Robert and Gullans, Michael and Lukin, Mikhail D. and Tanji-Suzuki, Haruka and Vuleti{\'c}, Vladan},
  title = {All-Optical Switch and Transistor Gated by One Stored Photon},
  journal = {Science},
  volume = {341},
  number = {6147},
  pages = {768--770},
  year = {2013},
  doi = {10.1126/science.1238169},
  publisher = {American Association for the Advancement of Science (AAAS)}
}

@article{Bechler2018PhotonAtomSwap,
  author = {Bechler, Orel and Borne, Adrien and Rosenblum, Serge and Guendelman, Gabriel and Mor, Ori Ezrah and Netser, Moran and Ohana, Tal and Aqua, Ziv and Drucker, Niv and Finkelstein, Ran and Lovsky, Yulia and Bruch, Rachel and Gurovich, Doron and Shafir, Ehud and Dayan, Barak},
  title = {A passive photon--atom qubit swap operation},
  journal = {Nature Physics},
  volume = {14},
  number = {10},
  pages = {996--1000},
  year = {2018},
  doi = {10.1038/s41567-018-0241-6},
  publisher = {Springer Nature}
}

@article{Wang2025ScalablePhotonicQT,
  author    = {Hui Wang and Timothy C. Ralph and Jelmer J. Renema and Chao-Yang Lu and Jian-Wei Pan},
  title     = {Scalable photonic quantum technologies},
  journal   = {Nature Materials},
  year      = {2025},
  volume     = {24},
  number     = {12},
  pages      = {1883--1897},
  doi        = {10.1038/s41563-025-02306-7}
}

@article{OBrien2009PhotonicQuantumTechnologies,
  author    = {Jeremy L. O'Brien and Akira Furusawa and Jelena Vu{\v{c}}kovi{\'c}},
  title     = {Photonic quantum technologies},
  journal   = {Nature Photonics},
  year      = {2009},
  volume    = {3},
  number    = {12},
  pages     = {687--695},
  doi       = {10.1038/nphoton.2009.229}
}

@article{chang2019microring,
  title={Microring resonators on a suspended membrane circuit for atom--light interactions},
  author={Chang, Tzu-Han and Fields, Brian M and Kim, May E and Hung, Chen-Lung},
  journal={Optica},
  volume={6},
  number={9},
  pages={1203--1210},
  year={2019},
  publisher={Optical Society of America},
    doi      = {10.1364/OPTICA.6.001203}
}

@article{wang2020integrated,
   title={Integrated photonic quantum technologies},
  author={Wang, Jianwei and Sciarrino, Fabio and Laing, Anthony and Thompson, Mark G},
  journal={Nature photonics},
  volume={14},
  number={5},
  pages={273--284},
  year={2020},
  publisher={Nature Publishing Group UK London},
  doi= {10.1038/s41566-019-0532-1},
}

@article{luo2023recent,
   title={Recent progress in quantum photonic chips for quantum communication and internet},
  author={Luo, Wei and Cao, Lin and Shi, Yuzhi and Wan, Lingxiao and Zhang, Hui and Li, Shuyi and Chen, Guanyu and Li, Yuan and Li, Sijin and Wang, Yunxiang and Sun, Shihai and Karim, Muhammad Faeyz and Cai, Hong and Kwek, Leong Chuan and Liu, Ai Qun},
  journal={Light: Science \& Applications},
  volume={12},
  number={1},
  pages={175},
  year={2023},
    doi={10.1038/s41377-023-01173-8},
    url={https://doi.org/10.1038/s41377-023-01173-8}
}

@article{corsetti2026integrated,
  title={Integrated-photonics-based systems for polarization-gradient cooling of trapped ions},
  author={Corsetti, Sabrina M and Hattori, Ashton and Clements, Ethan R and Knollmann, Felix W and Notaros, Milica and Swint, Reuel and Sneh, Tal and Callahan, Patrick T and West, Gavin N and Kharas, Dave and Mahony, Thomas and Bruzewicz, Colin D. and Sorace-Agaskar, Cheryl and McConnell, Robert and Chuang, Isaac L. and Chiaverini, John and Notaros, Jelena},
  journal={Light: Science \& Applications},
  volume={15},
  number={1},
  pages={57},
  year={2026},
  doi      = {10.1038/s41377-025-02094-4},
  publisher={Nature Publishing Group UK London}
}

@article{christen2025integrated,
  title={An integrated photonic engine for programmable atomic control},
  author={Christen, Ian and Propson, Thomas and Sutula, Madison and Sattari, Hamed and Choong, Gregory and Panuski, Christopher and Melville, Alexander and Mallek, Justin and Brabec, Cole and Hamilton, Scott and Dixon, P. Benjamin and Menssen, Adrian J. and Braje, Danielle and Ghadimi, Amir H. and Englund, Dirk},
  journal={Nature Communications},
  volume={16},
  number={1},
  pages={82},
  year={2025},
    doi      = {10.1038/s41467-024-55423-3},
  publisher={Nature Publishing Group UK London}
}

@article{blumenthal2024enabling,
  title={Enabling photonic integrated 3{D} magneto-optical traps for quantum sciences and applications},
  author={Blumenthal, Daniel J and Isichenko, Andrei and Chauhan, Nitesh},
  journal={Optica Quantum},
  volume={2},
  number={6},
  pages={444--457},
  year={2024},
    doi      = {10.1364/OPTICAQ.532260},
  publisher={Optica Publishing Group}
}

@article{goban2012demonstration,
  title = {Demonstration of a State-Insensitive, Compensated Nanofiber Trap},
  author = {Goban, A. and Choi, K. S. and Alton, D. J. and Ding, D. and Lacro\^ute, C. and Pototschnig, M. and Thiele, T. and Stern, N. P. and Kimble, H. J.},
  journal = {Phys. Rev. Lett.},
  volume = {109},
  issue = {3},
  pages = {033603},
  numpages = {5},
  year = {2012},
  month = {Jul},
  publisher = {American Physical Society},
  doi = {10.1103/PhysRevLett.109.033603},
  url = {https://link.aps.org/doi/10.1103/PhysRevLett.109.033603}
}

@article{vetsch2010optical,
  title = {Optical Interface Created by Laser-Cooled Atoms Trapped in the Evanescent Field Surrounding an Optical Nanofiber},
  author = {Vetsch, E. and Reitz, D. and Sagu\'e, G. and Schmidt, R. and Dawkins, S. T. and Rauschenbeutel, A.},
  journal = {Phys. Rev. Lett.},
  volume = {104},
  issue = {20},
  pages = {203603},
  numpages = {4},
  year = {2010},
  month = {May},
  publisher = {American Physical Society},
  doi = {10.1103/PhysRevLett.104.203603},
  url = {https://link.aps.org/doi/10.1103/PhysRevLett.104.203603}
}

@article{2026nearfieldheating,
      title={Limits of Stable Near-Field Probing in Nanophotonic Traps}, 
      author={Johannes Piotrowski and Constanze Bach and Nicolás Vera Paz and Philipp Schneeweiss and Arno Rauschenbeutel},
      year={2026},
      journal={arXiv: 2605.07798},
      url={https://arxiv.org/abs/2605.07798}, 
}

@article{hansen2026realization-142, 
  year    = {2026}, 
  title   = {Realization of waveguide many-body quantum optics}, 
  author  = {Hansen, Lena M and Henke, Clara and Hotter, Christoph and Sandberg, Oliver A D and Sandø, Thomas Wilkens and Angelopoulou, Vasiliki and Tiranov, Alexey and Møller, Christoffer B and Liu, Zhe and Midolo, Leonardo and Bart, Nikolai and Ludwig, Arne and Walther, Philip and Diepen, Cornelis J van and Lodahl, Peter and Sørensen, Anders Søndberg}, 
  journal = {arXiv: 2605.18525}, 
  year = {2026},
url={https://arxiv.org/abs/2605.18525}, 
}

@article{2023rmp_wQED,
  title = {Waveguide quantum electrodynamics: Collective radiance and photon-photon correlations},
  author = {Sheremet, Alexandra S. and Petrov, Mihail I. and Iorsh, Ivan V. and Poshakinskiy, Alexander V. and Poddubny, Alexander N.},
  journal = {Rev. Mod. Phys.},
  volume = {95},
  issue = {1},
  pages = {015002},
  numpages = {59},
  year = {2023},
  month = {Mar},
  publisher = {American Physical Society},
  doi = {10.1103/RevModPhys.95.015002},
  url = {https://link.aps.org/doi/10.1103/RevModPhys.95.015002}
}

@article{2016PRL_Jappel,
  title = {Coherent Backscattering of Light Off One-Dimensional Atomic Strings},
  author = {S\o{}rensen, H. L. and B\'eguin, J.-B. and Kluge, K. W. and Iakoupov, I. and S\o{}rensen, A. S. and M\"uller, J. H. and Polzik, E. S. and Appel, J.},
  journal = {Phys. Rev. Lett.},
  volume = {117},
  issue = {13},
  pages = {133604},
  numpages = {5},
  year = {2016},
  month = {Sep},
  publisher = {American Physical Society},
  doi = {10.1103/PhysRevLett.117.133604},
  url = {https://link.aps.org/doi/10.1103/PhysRevLett.117.133604}
}

@article{2016PRL_laurat,
  title = {Large Bragg Reflection from One-Dimensional Chains of Trapped Atoms Near a Nanoscale Waveguide},
  author = {Corzo, Neil V. and Gouraud, Baptiste and Chandra, Aveek and Goban, Akihisa and Sheremet, Alexandra S. and Kupriyanov, Dmitriy V. and Laurat, Julien},
  journal = {Phys. Rev. Lett.},
  volume = {117},
  issue = {13},
  pages = {133603},
  numpages = {6},
  year = {2016},
  month = {Sep},
  publisher = {American Physical Society},
  doi = {10.1103/PhysRevLett.117.133603},
  url = {https://link.aps.org/doi/10.1103/PhysRevLett.117.133603}
}

@article{1978correlation,
doi = {10.1088/0305-4470/11/5/007},
url = {https://doi.org/10.1088/0305-4470/11/5/007},
year = {1978},
month = {may},
publisher = {},
volume = {11},
number = {5},
pages = {L121},
author = {H J Carmichael and P Drummond and P Meystre and D F Walls},
title = {Intensity correlations in resonance fluorescence with atomic number fluctuations},
journal = {Journal of Physics A: Mathematical and General},
}

@article{1977PRL_photonantubunching,
  title = {Photon Antibunching in Resonance Fluorescence},
  author = {Kimble, H. J. and Dagenais, M. and Mandel, L.},
  journal = {Phys. Rev. Lett.},
  volume = {39},
  issue = {11},
  pages = {691--695},
  numpages = {0},
  year = {1977},
  month = {Sep},
  publisher = {American Physical Society},
  doi = {10.1103/PhysRevLett.39.691},
  url = {https://link.aps.org/doi/10.1103/PhysRevLett.39.691}
}

@article{2025PRApplied_Leixu,
  title = {Dynamics of single atoms in optical tweezers near a chip's surface},
  author = {Xu, Lei and Wang, Ling-Xiao and Chen, Guang-Jie and Wang, Zhu-Bo and Xu, Xin-Biao and Guo, Guang-Can and Zou, Chang-Ling and Xiang, Guo-Yong},
  journal = {Phys. Rev. Appl.},
  volume = {24},
  issue = {2},
  pages = {024002},
  numpages = {11},
  year = {2025},
  month = {Aug},
  publisher = {American Physical Society},
  doi = {10.1103/rd53-4w5w},
  url = {https://link.aps.org/doi/10.1103/rd53-4w5w}
}

\section*{Acknowledgement}
We thank S. Bhave, T.-H. Chang, and Y. Tian for discussions on the photonic integrated circuits. We acknowledge Hikaru Tamura and Lyuhang Wu for the help on the initial stage of this experiment and acknowledge S. J. V. Satya for technical assistance during the preparation of the manuscript. This work was supported by the AFOSR (Grant NO. FA9550-22-1-0031 \& FA955-026-1-B086), the ONR (Grant NO. N000142412184), and the Gordon and Betty Moore Foundation (grant DOI: 10.37807/GBMF13795). D. D. acknowledges the Rolf Scharenberg Graduate Summer Research Fellowship.

\newpage
\section*{Methods}

\renewcommand{\figurename}{\textbf{Fig.}}

\renewcommand{\thefigure}{S\arabic{figure}}
\setcounter{figure}{0}

\subsection{Cooling, transporting, and probing atoms in an optical conveyor belt}

\begin{figure}[b]
\centering
\includegraphics[width=0.9\columnwidth]{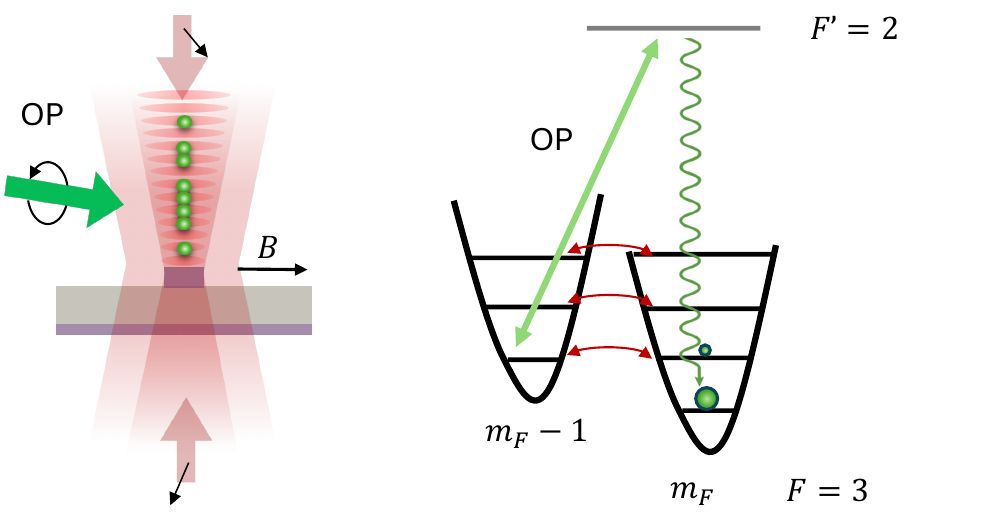}
\caption{\textbf{Schematics for the dRSC.} A circularly polarized optical pumping beam (OP) illuminates the PIC at a shallow $5^\circ$ grazing angle, driving the $\sigma^+$ transition. Slightly misaligned polarization (illustrated by black arrows) in the top and bottom beams create the degenerate Raman coupling (red curved arrows) between the adjacent magnetic levels in exchange of one vibrational quanta. In the Lamb-Dicke regime, the optical pumping process cools the atoms to lower vibrational states. 
}
\label{fig:drsc}
\end{figure}

The experimental setup for initial magneto-optical trapping and optical guiding on a PIC has been discussed in detail in Ref.~\cite{Zhou2023CouplingGuiding}. Following ramping on the optical tweezer from the top, we perform degenerate Raman sideband cooling (dRSC)~\cite{Vuletic1998DegenerateDensities} to load the guided atoms into an initially static optical lattice, as schematically shown in Fig.~\ref{fig:drsc}. The optically guided atoms are first spin-polarized in the $\ket{F=3}$ hyperfine ground state. We apply the spin-motion coupling for dRSC by deliberately misaligning the polarization angle ($\leq 5^\circ$) between the top and bottom beams, thereby introducing a small vector potential to drive the desired Raman transitions. We tune the background magnetic field so that the Zeeman splitting between magnetic sublevels coincide with the vibrational level spacing of the lattice trap. An external optical pumping beam drives the $\sigma^{+}$ transition for $\ket{F=3} \rightarrow \ket{F^{\prime}=2}$. The dRSC operates in the Lamb-Dicke regime, where the motional excitation is suppressed during the optical pumping process. As illustrated in Fig.~\ref{fig:drsc}, one sample cooling cycle starts with a degenerate Raman transition, reducing one vibrational energy quanta in exchange of $\Delta m_F = -1$, which is followed by optical pumping and spontaneous emission back to the original magnetic level. This sequence efficiently cools the atoms by pumping them into the $\ket{F=3,m_F=3}$ dark state, which has the lowest magnetic and trap energies. We estimate the temperature of the trapped atoms ($\lesssim 20~\mu$K) by performing a two-photon Raman spectroscopy to measure the transferred population in the red and blue sidebands. 

After dRSC, the guided atoms are fully localized in the conveyor belt. We then optically pump the trapped atoms back to the $\ket{F=4,m_F=4}$ hyperfine ground state, followed by ramping to a non-zero, static detuning $\Delta \nu$ between the top and bottom beams to initiate the conveyor-belt transport. During each transport, we continuously monitor the probe transmission through the bus waveguide using an avalanche photodiode (APD) and a time tagger. After a sufficiently long time to deplete all trapped atoms, we record the background transmission and restart the experimental cycle.

\subsection{Feedback on the conveyor-belt transport}
\begin{figure}[!ht]
\centering
\includegraphics[width=1.0\columnwidth]{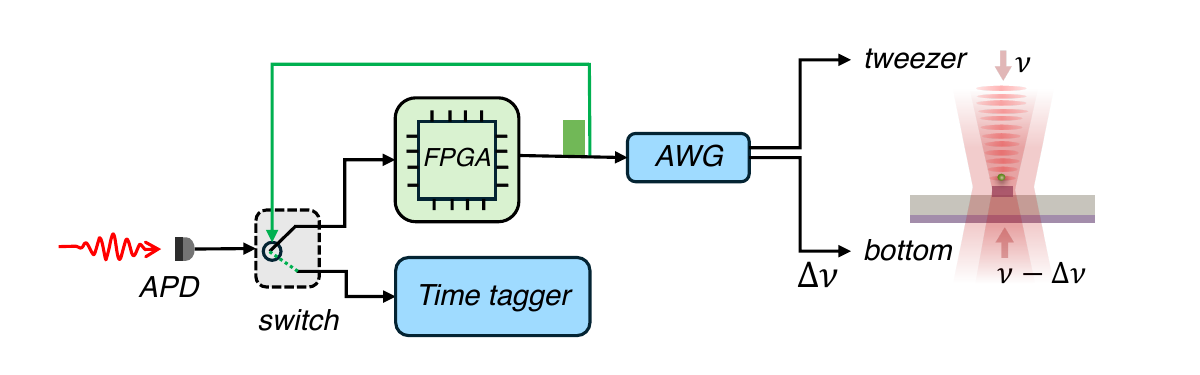}
\caption{\textbf{Schematic for the feedback.}
}
\label{fig:fig5}
\end{figure}

In Fig.~\ref{fig:fig4}, we implement real-time feedback on the conveyor-belt transport using a simple FPGA (Red Pitaya). The FPGA keeps track of the accumulated APD counts in a 10~$\mu$s running window. Upon reaching a threshold $C_\mathrm{th}=10$, the FPGA triggers the optical tweezer power to ramp up by four times in 5~$\mu$s to localize the detected atoms in a stationary trap near the surface, while simultaneously triggering the AWG to ramp down $\Delta \nu$ in 50~$\mu$s to completely halt the conveyor belt; see Fig.~\ref{fig:fig5}. We have engineered a smooth RF profile such that all atoms remain stably trapped during the frequency ramp-down without heating them out of the optical lattice. During and after the trap transfer, probe transmission is continuously registered by the APD and the time tagger, as shown in Fig.\ref{fig:fig4}(f-g).

\subsection{Lifetime measurement}
\begin{figure}[!ht]
\centering
\includegraphics[width=0.8\columnwidth]{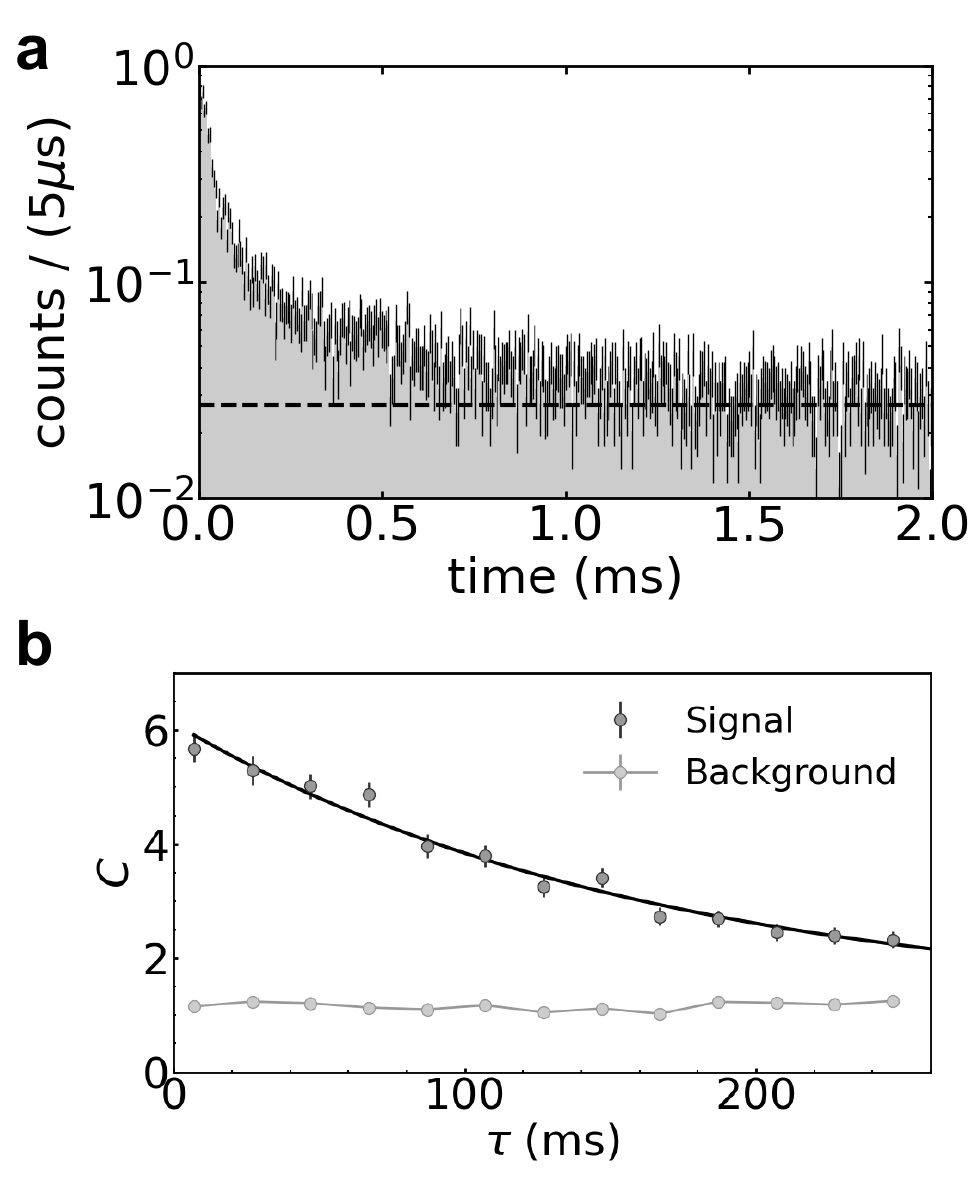}
\caption{\textbf{Resonance fluorescence and lifetime measurement.}
\textbf{a}, Time dependence of resonance fluorescence counts collected through the microring. \textbf{b}, Fluorescence counts integrated over the first $200\mu$s of measurement as a function of the hold time $\tau$. The black solid line shows an exponential fit, giving a $1/e$-lifetime of $ 162(8)$~ms.
}
\label{fig:fig6}
\end{figure}

In a separate experiment, we determine the lifetime of trapped atoms near the microring surface. Following the dRSC in the stationary optical conveyor belt and a variable hold time $\tau$ in the ground state, we illuminate the atoms externally using a weak resonant beam and collect the fluorescence using the microring circuit. Only those atoms localized within the evanescent field region can be detected. As shown in Fig.~\ref{fig:fig6}(a), the fluorescence signal decays in $\sim200~\mu$s due to resonant heating, which is consistent with the time scale measured in Fig.~\ref{fig:fig4}(g). Figure \ref{fig:fig6}(b) shows fluorescence counts versus hold time $\tau$, indicating a trap lifetime of $162(8)~$ms similar to that of trapped atomic ensembles demonstrated in Ref.~\cite{2024PRX_trappedatoms}.

\subsection{Single atom detection}
To confirm that the site occupancy in the optical conveyor belt is indeed close to unity, we further analyze the photon correlations in the resonance fluorescence using a Hanbury-Brown–Twiss (HBT) setup. We search for the photon antibunching in the resonance fluorescence \cite{1977PRL_photonantubunching}, which is expected from single atom emissions. We first analyze the photon correlations by evaluating
\begin{equation}
\xi(\tau)= \left<\frac{ \overline{I_1(t)I_2(t+\tau)}}{\overline{I_1(t)} \cdot \overline{I_2(t)}}\right>\,,\label{eq_xi}
\end{equation}
where $I_{i}$ denotes the count trace of detector $i=\{1,2 \}$ and $\bar{.}$ ($\braket{\cdot}$) indicates the time averaging over $200\,\mu\rm{s}$ of signal (ensemble averaging over repeated experiments). 

\begin{figure}[t]
\centering
\includegraphics[width=0.8\columnwidth]{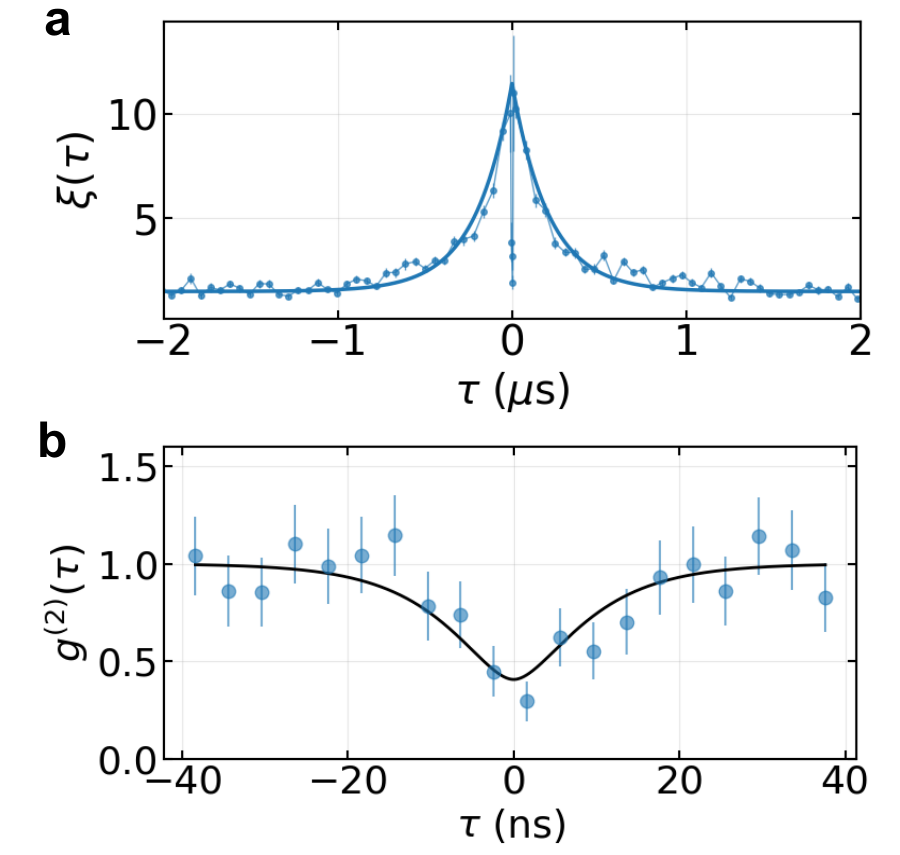}
\caption{\textbf{Photon correlation measurements for the resonance fluorescence. } \textbf{a}, Cross-correlation $\xi(\tau)$ of photon counts from two detectors. 
\textbf{b}, correlation function $g^{(2)}(\tau)$ of the resonance photons when trapped atoms are in the strong coupling regime. The solid line shows a theoretical fit which gives a mean atom number $\bar{N}=1.2(2)$ and decay rate $\gamma=2\pi\times 52(5)~$MHz. 
}
\label{fig:figg2}
\end{figure}

Figure~\ref{fig:figg2}(a) shows photon bunching in a time scale of $0\ll|\tau|<1~\mu$s, which is related to the atomic motion during the resonance driving. At much shorter time delays, we observe great reduction of photon correlation which appears as a sharp dip at $\xi(0)$. However, the observed $\xi(0)\gtrsim 1$ is due to fluctuations in the atom number and position.

To clearly resolve the antibunching signal, we analyze the recorded photon count traces in select time windows ($2\,\mu\rm{s}$ wide) when the count rate is higher than 2 counts per $1.6\,\mu\rm{s}$, and similarly calculate $g^{(2)}(\tau)$ in this time window using Eq.~\eqref{eq_xi}. This procedure analyzes resonance fluorescence when the trapped atoms are present in the strong coupling region. 
Figure~\ref{fig:figg2}(b) shows the resulting correlation $g^{(2)}(\tau)$ and reveals strong antibunching with $g^{(2)}(0)<0.5$ below the classical limit, indicating the resonance fluorescence is primarily emitted by single atoms. On the other hand, the time scale for observing the antibunching is $<\Gamma^{-1}\approx30~$ns, suggesting strongly Purcell-enhanced emissions.  
 
We compare the measurement with theory to estimate the mean number of trapped atoms. The analysis of normalized correlation function $g^{(2)}(\tau)$ with a fixed atom number $N$ can be found in Ref.~\cite{1978correlation}, which reads 
\begin{equation}
g_N^{(2)}(\tau)=\frac{G_N^{(2)}(\tau)}{\left|G_N^{(1)}(0)\right|^2} \label{eq:gn}
\end{equation}
with
\begin{equation}
\begin{aligned}
G_N^{(2)}(\tau)=N G_A^{(2)}(\tau)+ N(N-1)\left[I^2+\left|G_A^{(1)}(\tau) \right|^2
\right]\\
+I_\epsilon^2+2 I N I_\epsilon+
2\,\mathrm{Re}\!\left(G_A^{(1)}(\tau)\right) NI_\epsilon.
\end{aligned}
\end{equation}
and
\begin{equation}
G_N^{(1)}(0)= NI + I_\epsilon .
\end{equation}
Here $G_A^{(1),(2)}$ are single-atom first- and second-order correlation functions, as defined in Ref.~\cite{1978correlation}, $I$ is the intensity from a single atom, and $I_\epsilon$ is the background. At zero background $I_\epsilon=0$, $g_N^{(2)}(0)>0.5$ for $N\geq2$ and $g_1^{(2)}(0)=0$. Under this model, any measured nonzero $g^{(2)}(0)$ will be attributed to contributions from atom number above one. 

We implement a fit using the Poisson average of $g_N^{(2)}(\tau)$ in Eq.~\eqref{eq:gn} in the limit of weak driving and background-free detection. From the fit as shown in Fig.~\ref{fig:figg2}(b), we obtain mean atom number $\bar{N}=1.2(2)$ and decay rate $\gamma=2\pi\times 52(5)~$MHz. The mean atom number is in very good agreement with the averaged population determined in Fig.~\ref{fig:fig3}(a). The fitted decay rate $\gamma\sim 10\Gamma$ confirms that large Purcell enhancement and single atom-photon strong coupling is reached.

\subsection{Numerical simulation}
To evaluate the conveyor-belt trap potential, we compute the electric field patterns of the top and bottom-illuminating dipole beams using finite-difference time domain (FDTD) simulations (Lumerical FDTD). We then superimpose the counter-propagating electric fields with a time-varying relative phase and a random initial phase offset to calculate the time-dependent conveyor-belt potential $U_\mathrm{lat}$. The evanescent field of the resonator mode at wavelength $\lambda_\mathrm{b}$ is computed using a finite element method (COMSOL Multiphysics), which forms a repulsive potential $U_\mathrm{b}$ near the surface of the microring. In addition, we evaluate the Casimir-Polder potential using FDTD calculations~\cite{hung2013trapped}, fitted with an empirical form $U_\mathrm{CP}(z)=C_{4,\rm{eff}}/[\tilde{z}^3(\tilde{z}+\tilde{\lambda})]$ for $|y|<w/2$~\cite{2024PRX_trappedatoms}, where $\tilde{\lambda}=\lambda/2\pi$, $\tilde{z}=z-z_\mathrm{0,eff}$, $C_{4,\rm{eff}}/h=-165.36~$Hz$\cdot\mu$m$^4$, $z_\mathrm{0,eff}=3.7~$nm, and $w=950~$nm is the width of the waveguide. The total trap potential is $U=U_\mathrm{lat} + U_\mathrm{b} + U_\mathrm{CP}$, as plotted in Figs.~\ref{fig:fig2}(a) and \ref{fig:fig4}(d). For further details on dipole traps, light shifts, and $U_\mathrm{CP}$ calculations, see also Refs.~\cite{Zhou2023CouplingGuiding,2024PRX_trappedatoms}.

To simulate the dynamics of probe transmission, we perform Monte Carlo simulations to sample atomic trajectories in a moving conveyor belt. As the experiment was performed above a straight part of the microring waveguide, we only consider the atomic motion in the $y$-$z$ plane defined in Fig.~\ref{fig:fig1}(b). We sample the initial condition of a trapped atom, localized near a site center, using the Maxwell-Boltzmann distribution. We then evolve the atomic motion in a moving lattice using the fourth-order Runge Kutta method. To prevent erratic trajectories near the surface, the simulation is terminated as long as the atom enters $z\leq10~$nm.

Given an atomic trajectory $\vec{r}(t)=(y(t),z(t))$, we evaluate the atom-photon coupling rate $g(\vec{r})$ using an empirical function $g(\vec{r})\approx g_0e^{-z/\lambda_z}\cos(y/l_y)e^{-y^2/\lambda_y^2}$ obtained by fitting the mode field distribution. Here $g_0\approx 2\pi\times490~$MHz, $\lambda_z\approx 88~$nm, $l_y\approx 540~$nm and $\lambda_y\approx 930~$nm. The cooperativity parameter $C=4g^2/(\kappa\Gamma)$ is evaluated accordingly, as shown in Figs.~\ref{fig:fig2}(b) and \ref{fig:fig4}(e). 

To evaluate the transmission of the probe field due to atom-induced transparency, we assume the resonator photon remains in the steady state since $\kappa \gg g$. We calculate the transmission spectrum
\begin{equation}
    T_1 = \left|1- \frac{2\kappa_e \tilde{\eta}}{\tilde{\kappa}}\frac{1}{1+\tilde{\eta}\tilde{C}} \right|^2, \label{eq:T1}
\end{equation}
where $\kappa_\mathrm{e}$ is the bus waveguide coupling rate, $\tilde{\eta} = 1/(1+4\beta^2/\tilde{\kappa}^2)$ is a reduction factor due to back-scattering in the microring, $\tilde{\kappa}=\kappa+2i\delta$, $\tilde{C} = 4g^2/(\tilde{\kappa}\tilde{\Gamma})$, $\tilde{\Gamma} \approx \Gamma + 2i\delta'$, $\delta = \omega_\mathrm{c} -\omega $ is the probe detuning from the cavity resonance frequency $\omega_\mathrm{c}=\omega_\mathrm{a}$, $\delta' = \omega_\mathrm{a} + \delta_\mathrm{LS} -\omega$ is the detuning from the atomic resonance, and $\delta_\mathrm{LS}$ is the calculated position-dependent differential light shift due to the blue-detuned evanescent field and the Casimir-Polder surface interaction. From fitting the transmission spectrum as shown in Fig.~\ref{fig:fig1}(c), we have determined $(\kappa, \kappa_\mathrm{e}, \beta) = 2\pi \times (1.71, 0.83, 0.51)~$GHz. 

When $n>1$ atoms are trapped in a single lattice site, they collectively interact with the resonator photon, further enhancing the transmission. We replace the modified cooperativity parameter $\tilde{C}$ in the steady-state calculation Eq.~\eqref{eq:T1} with
\begin{equation}
    \tilde{C}_n =  \sum_{j=1}^n \tilde{C}(\vec{r}_j)\,,
\end{equation}
where $\vec{r}_j$ is the instantaneous position of the $j$-th atom, $\tilde{C}(\vec{r}_j) = 4g^2(\vec{r}_j)/(\tilde{\kappa}\tilde{\Gamma}_j)$, and $\tilde{\Gamma}_j = \Gamma + 2i \delta'(\vec{r}_j)$ considers the position-dependent light shift of the $j$-th atom. The resulting transmission $T_n$ is plotted in Fig.~\ref{fig:fig2}.

\begin{figure}[b!]
\centering
\includegraphics[width=0.8\columnwidth]{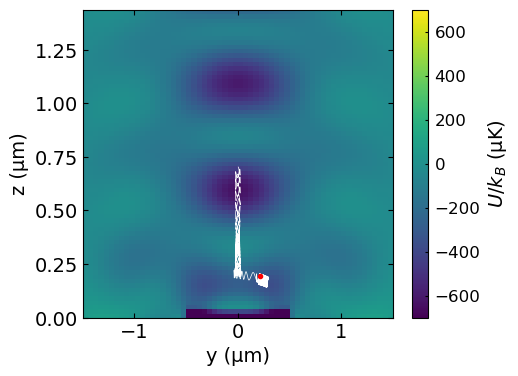}
\caption{\textbf{Simulated trajectory with feedback.}
The trajectory (white line) is plotted from the initial time ($t=0$) till around 180$~\mu$s after the `trigger,' which occurs at $t\approx151~\mu$s. The trajectory overlays the final stationary potential cross-section $U(y,z)$, and the final atomic position is marked by the red circle. The trajectory localizes to the right side lobe immediately after the tweezer power is ramped up. The waveguide surface spans from $(y,z)=(-475,0)$~nm to (475, 0)~nm.
}
\label{fig:trajectory}
\end{figure}

In Fig.~\ref{fig:fig4}, we have additionally simulated the atomic trajectories and the atom-photon coupling in the event of triggering. We simulate the ramping of the top tweezer power and the stopping of the conveyor belt once an atom reaches a threshold position $z_\mathrm{th}=200~$nm above the microring surface, which roughly corresponds to the threshold condition of $C_\mathrm{th}=1$ per microsecond. Upon ramping up the tweezer power, the simulated trajectory becomes localized in one of the two stationary `side lobes' located at $(y,z) \approx (\pm 245, 170)~$nm above the microring surface, as shown in Fig.~\ref{fig:fig4} (d). These traps result from the interference between the tweezer and its reflection in the near-field of the rectangular waveguide structure. The blue-detuned evanescent field further provides a repulsive barrier near the surface. Our choice of slow conveyor-belt motion of $2.8~$nm/$\mu$s and the fast ramp-up time of the tweezer power enables nearly 100\% transfer observed in the numerical simulations using our experiment parameters. See Fig.~\ref{fig:trajectory} for one sample trajectory. Figure~\ref{fig:fig4}(e) plots the atom-photon coupling rate $g$ right after the trigger, indicating a stable trajectory within the stationary trap. We have calculated the time-averaged transmission $\bar{T}_n \approx 0.20$ and 0.35 for $n=1$ and 2 atoms loaded in the stationary trap. Note that at this position the calculated light shift~\cite{Zhou2023CouplingGuiding} $\delta_\mathrm{LS}\approx -2\pi\times 6~$MHz is primarily due to the blue-detuned evanescent field potential, which reduces the probe transmission signal at $\delta=0$ by around 1.9 times compared with the transmission at $\delta'=0$. 

\subsection{Determination of $n$-atom transfer probability}
We determine the success rate for trapping $n$-atoms by analyzing the histogram of the APD counts as shown in Fig.~\ref{fig:fig4}(f). 
We fit the normalized occurrence $O$ with a composite Gaussian model
\begin{equation}
    O(I)=\frac{1}{\sqrt{\pi}}\left[\frac{P_0}{w_{bg}}e^{-\frac{(I-I_{bg})^2}{w_{bg}^2}}+\sum_{n=1}^{2}\frac{P_n}{w\sqrt{I_n}}e^{-\frac{(I-I_n)^2}{w^2 I_n}}\right] \,,
\end{equation}
where $I$ is the integrated counts, $P_n$ is the probability of having $n=0,1,2$ atoms in the trap. We first extract $(I_\mathrm{bg}, w_\mathrm{bg})=(15, 6)$ using a single Gaussian fit (by setting $P_1=P_2=0$) to the background-only histogram. For fitting the signal with trapped atoms, we assume $I_n=sI_\mathrm{bg}\bar{T}_n/\bar{T}_0$ while fixing $I_\mathrm{bg}$ and using the calculated $\bar{T}_n/\bar{T}_0=2.35$ and 4.12 for $n=1$ and 2 atoms ($\bar{T}_0=0.085$ is the background). We allow the scale factor $s$ and the excessive noise factor $w$ as fit parameters to account for the experiment imperfections. In addition, we set the constraint $P_2=1-P_0-P_1$ in the composite Gaussian fit. 

Our fit determines $(P_0, P_1, w, s)=(0\pm0.031, 0.818\pm0.051, 1.86\pm0.22, 0.92\pm0.03)$, giving $P_2 = 0.18\pm 0.06$. The result suggests $\gtrsim 97\%$ success rate for trap transfer, with a probability $P_1\approx 82(5)\%$ ($P_2\approx 18(6)\%$) for trapping single atom (two atoms). The scale factor $s\approx 0.92$ is slightly below unity, which may be due to the inaccuracy in our simulated final trap position and the calculated differential light shift reducing the atom-induced transmission. While these can be fully calibrated, in the future experiments the light shift can be eliminated by adopting a magic wavelength at $\lambda_\mathrm{b}\approx 794~$nm~\cite{chang2019microring}. If we attribute the scale factor to the uncertainty of trap position, which leads to the reduction of atom-photon coupling, we arrive at $g/2\pi=48(1)~$MHz and $C\approx 1.04(4)$.

\end{document}